Laser-Induced Graphitisation of Diamond Under 30 *fs* Laser Pulse Irradiation

AUTHORS

Bakhtiar Ali [a, b], Han Xu [b, c], Dashavir Chetty [b, c], Robert T. Sang [b, c], Igor V. Litvinyuk [b, c], Maksym Rybachuk [a, b]*

AFFILIATIONS

[a] School of Engineering and Built Environment, Griffith University, 170 Kessels Rd., Nathan QLD 4111, AUSTRALIA

[b] Centre for Quantum Dynamics and Australian Attosecond Science Facility, Griffith University, Science Road, Nathan, QLD, 4111, AUSTRALIA

[c] School of Environment and Science, Griffith University, Nathan QLD 4111, AUSTRALIA

* Corresponding author:

m.rybachuk@griffith.edu.au

https://orcid.org/0000-0002-5313-9204

TOC Graphic

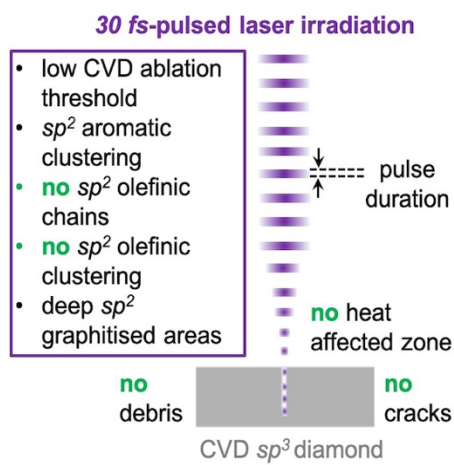

ABSTRACT

The degree of laser-induced graphitisation from a $sp^3$-bonded to a $sp^2$-bonded carbon fraction in a single crystal chemical vapour deposited (CVD) diamond under a varying fluence of an ultrashort pulsed laser (30 $fs$, 800 nm, 1 kHz) irradiation has been studied. The tetrahedral CVD $sp^3$-phase was found to transition to primarily an $sp^2$-aromatic crystalline graphitic fraction below the critical fluence of 3.9 J/cm$^2$, above which predominantly an amorphous carbon was formed. A fractional increase of fluence from 3.3 J/cm$^2$ to 3.9 J/cm$^2$ (~ 20 %) resulted in a substantial (~ three-fold) increased depth of the $sp^2$-graphitised areas owing to the non-linear interactions associated with an $fs$-laser irradiation. Additionally, formation of C=O carbonyl group was observed below the critical threshold fluence; the C=O cleavage occurred gradually with the increase of irradiation fluence of 30 $fs$ laser light. The implications for these findings on enhancement of $fs$-driven processing of diamond are discussed.

Highly localised and geometrically confined structural transformations in diamond with intrinsic $sp^3$ diamond's tetrahedral phase converted into $sp^2$-rich graphitic fraction employing a laser light have shown to significantly enhance light absorption in the irradiated areas and, to open a new avenue for the emerging opto-photonic applications, including miniaturised thermionic solar cells[1], broad-beam light detectors[2-3] and most recently, robust broad-beam polarization filters for infra-red (IR) applications[4]. These applications, however, may only become technologically accomplishable once the thermal stresses associated with laser processing and the formation of $sp^2$ fraction on the surface or in the bulk of the diamond crystal, are minimised while remaining spatially confined. Additionally, owing to a close lattice match between the diamond's $sp^3$ tetrahedral fraction and the planar two-dimensional (2D) graphene and graphite structures, direct fabrication of 'graphene-on-diamond' heterostructures by means of an *in situ* $sp^3$-to-$sp^2$ conversion enables a development of a variety of robust ultra-wide bandgap 'all carbon' opto-electronic devices in a single-step fabrication process, without resorting to processes that normally require high-temperature annealing with a metal catalyst.[5-6]

Conventional nano- (*ns*-) and pico- (*ps*-) second pulsed lasers transform diamond into graphite by means of a localised phonon-induced graphitization and, owing to their long laser-lattice interactions, primarily an amorphous $sp^2$ phase is formed in the heat affected zone (HAZ) marked by the appearance of thermally-induced cracks.[7] The ultrashort femtosecond (*fs*) laser pulses generate minimal to no HAZ and produce no HAZ-associated localised cracking as the *fs* pulse ends before the excited electrons are able to transfer their energy to the diamond lattice as the characteristic time of electron-lattice interaction is a few *ps*, at least an order of magnitude longer than the pulse duration.[8-9] However, it is known that pulses as short as 100 *fs* still thermalise electrons and generate hot ions and phonons leading to thermal quenching and $sp^3$ lattice relaxation in diamond.[7, 10-12] It has been postulated that pulses shorter than 50 *fs* display primarily non-thermal characteristics as photo-ablation and structural re-organisation is predominantly driven by photo-ionization of electrons.[10, 13] The sub-50 *fs* regime still remains largely unexplored for processing of diamond.[7] The volumetric processing efficiency, that is the *sp³* into *sp²* phase conversion, followed by a subtractive *sp²* fraction removal, remains low for all pulse widths and the lengths of the laser pulse being used, as the thermal diffusion is primarily limited to the focal volume.[14] Processing diamond above its graphitisation threshold with *ns*- and/or *ps*- pulses is associated with a higher volume/mass removal but also with a significant and unavoidable formation of HAZ as longer pulses generate plasma plumes of higher ion density and electron temperature.[9, 15] *Fs*-pulses and, in particular, the sub-50 *fs* pulses are able to deliver a fluence well above the diamond's graphitization threshold (0.3 J/cm$^2$)[16] at or near the ablation threshold (~ 3.0 – 4.0 J/cm$^2$)[17-18] and, notionally, without any thermal damage. Simulations suggest that under *ns*- or longer laser pulse durations, graphitization propagates vertically affecting the bulk of the focal volume, leading to the formation of largely amorphous (*a*-) *sp³*:*sp²* interfaces following the laser treatment[19-20]. By contrast, with *fs*- laser pulses, graphitization of diamond occurs fully layer by layer in a *pseudo* 'peel off' process, resulting in the formation of a clean diamond surface after the ablation that is highly desirable for applied applications.

In this work, we present a preliminary experimental study on graphitisation of diamond employing an ultra-short laser pulse duration and demonstrate that a 30 *fs* photo-ablation of a single crystal CVD below its critical ablation threshold results primarily in a formation of *sp²*-aromatic graphitic fraction with a minimal amorphization, and in an instance when the fluence near or at the ablation threshold is used, a substantial increase (~ three-fold) of the formed *sp²*-graphitised layer on the CVD is observed. Fractional formation of C=O carbonyl group was also observed within the *sp²* graphitised regions under all irradiation condition. Our findings demonstrate a potential for the successful use of ultra-short duration laser pulses for processing of CVD and its derivatives (*i.e.,* natural diamond, nano- and micro-diamonds, bucky-diamond particles) for fabrication of diamond systems with previously unattainable composition and properties and, *fs* laser-assisted plating and sintering applications.

An 800 nm $Ti^{3+}$:sapphire laser system (FEMTOPOWER™ compact™ PRO HE, FEMTOLASERS Produktions GmbH, Austria) delivering a linearly polarized Gaussian beam with a 30 *fs* pulse duration at 1 kHz repetition rate was used to irradiate an *n*-type <100> Ib CVD diamond samples (3×3×1 mm) with ~200 ppm of nitrogen content (Chenguang Machinery & Electric Equipment Co. Ltd., Hunan, China). The source pulse energy of 0.8 mJ was attenuated to 5.4 *μ*J, 3.1 *μ*J, 2.6 *μ*J and 1.7 *μ*J using pellicle beam splitters producing a laser peak fluence of 6.8 J/cm$^2$, 3.9 J/cm$^2$, 3.3 J/cm$^2$ and 2.2 J/cm$^2$ and, a peak intensity of 2.3 x$10^{14}$ W/cm$^2$, 1.3 x$10^{14}$ W/cm$^2$, 1.1 x$10^{14}$ W/cm$^2$ and 0.73 x$10^{14}$ W/cm$^2$, respectively. An 90° off-axis parabolic silver mirror and a CMOS camera beam profiler (BC207VIS(/M), Thorlabs Inc., USA) with Beam™ software package were used to precisely focus the *fs*-laser beam over the CVD diamond sample surface. The focal spot size of the beam was 10 *μ*m at 1/e of the maximum intensity with an $M^2$ of 1.3 for all experiments. The samples positioned normal to the beam were irradiated at a scanning speed of 15 *m*m/s, 667 pulses per irradiation site. No self-phase modulation and/or filamentation in front of the focal point has been observed during the experiments. Unpolarised 514 nm $Ar^+$ ion laser (Renishaw inVia™, UK) was used to obtain micro-Raman spectra at 293 K from the laser-irradiated samples at 1 cm$^{-1}$ resolution and using 0.1 mW power over an extended acquisition time to minimise possible thermal damage to CVD samples. Fully symmetric Gaussians were used to re-constitute Raman bands following a linear photoluminescence (PL) background subtraction employing a non-liner least squares fitting procedure [21]; the coefficients of determination, $R^2$, of >0.99 were obtained for all fits. Sample surface topography was studied using optical (NIKON Eclipse E-200, Japan) and scanning electron microscopy (SEM) (JSM-6510, JEOL Ltd., Japan).

Figure 1(a) displays Raman spectra obtained from laser-induced graphitized tracks *ca.* 10 $\mu$m wide on CVD at varying fluence, including the *sp$^3$* tetrahedral diamond mode at ~1332 cm$^{-1}$ corresponding to the triply degenerate optical phonon of core T$_{2g}$ symmetry near the centre of the Brillouin zone with a sharp line with the half width at half maximum (HWHM) of ~6 cm$^{-1}$, the core *sp$^2$ D* and *G* modes at *ca.* 1350 cm$^{-1}$ and *ca.* 1582 cm$^{-1}$ corresponding to Raman active A$_{1g}$ and E$_{2g}$ modes, respectively, and a carbonyl group (-C=O) mode at *ca.* 1710 cm$^{-1}$.[22-25] The spectra display a pronounced PL scattering background associated with hydrogenation.[26-29] The PL gradient increased with increasing fluence, whereas the ~1332 cm$^{-1}$ diamond mode gradually receded and almost completely resolved at 6.8 J/cm$^2$ (blue spectrum), indicating significant graphitisation of irradiated areas on CVD.[30]

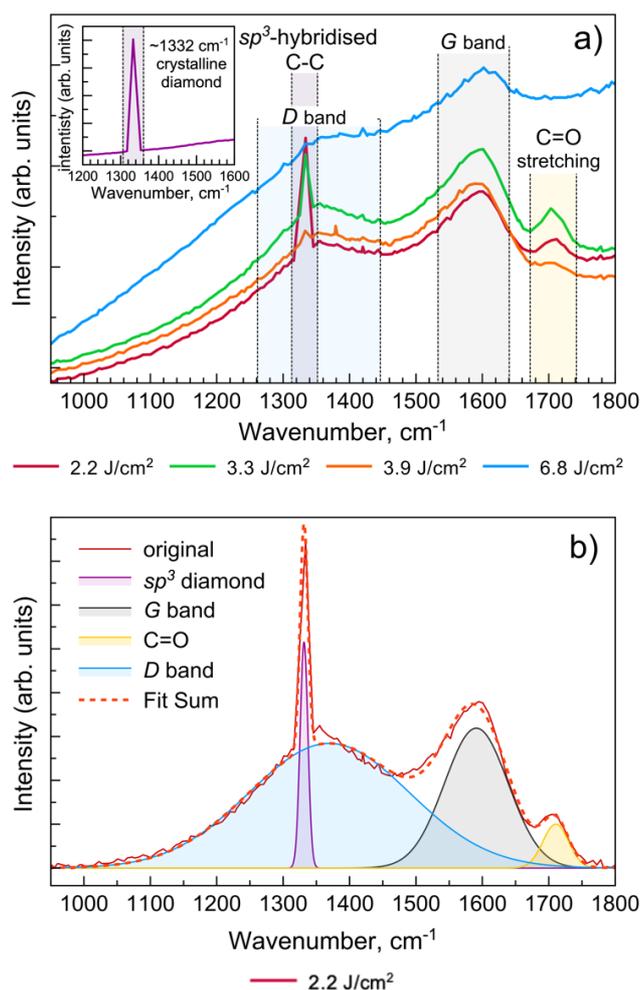

Fig. 1a) 514 nm Raman spectra of 30 *fs*-laser irradiated CVD at varying fluence, b) example of a deconvoluted Raman spectra of CVD irradiated at 2.2 J/cm² fluence showing the constituent *sp³* (1332 cm⁻¹) diamond band, and *D*, *G* and C=O bands.

Fig. 1b) shows deconvoluted Raman spectra of CVD irradiated at 2.2 J/cm² fluence. The constituent *sp³* tetrahedral diamond mode at ~1332 cm⁻¹, and *D*, *G* and C=O bands are clearly visible. Detailed analysis of Raman spectra, as shown in Fig, 2a-c, revealed two distinct spectral characteristics of *fs*-pulsed laser irradiated CVD samples as shown by the evolution of relative ratios of the *D* to the *G* peak, *I(D)/I(G)* (Fig. 2a), their respective HWHM values (Fig. 2b) and, relative positions of the *D*, $D_\Gamma$ and the *G*, $G_\Gamma$ peaks (Fig. 2c) at increasing fluence. Notably, the *I(D)/I(G)* ratio increased from 0.89 to 1.05, whereas the effective crystallite size, $L_a$, measured in nm, which was initially defined by Tuinstra and Koenig[27] and given by Cançado *et. al.*[31] for 514 nm laser excitation, *E* (*i.e.,* 2.41 eV), using the Eq. (1)

$$L_a = \frac{560}{E^4} \left(\frac{I(D)}{I(G)}\right)^{-1}, \tag{1}$$

was found to decrease from approximately 19 nm to 16 nm at increasing fluence from 2.2 J/cm² to 3.9 J/cm² (see Fig. 2a), indicating increasing *sp²* aromatic clustering near the K and M points of the Brillouin zone and corresponding the overall reduction of crystallinity.[23,32]

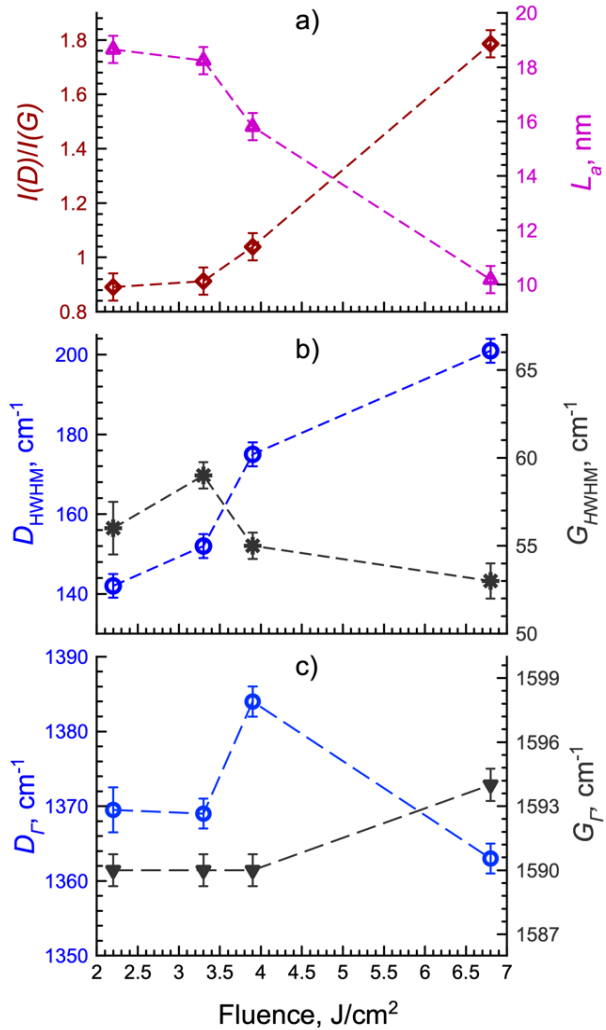

Fig. 2. Characteristic features of Raman spectra of 30 *fs*-laser irradiated CVD, including a) *I(D)/I(G)* peak ratios and effective crystallite size, $L_a$, b) HWHM values for the *D* and the *G* peaks and, c) positions of the *D* and *G* peaks at varying fluence.

These observations are commensurate with those reported by Girolami *et. al.*[32] for 100-*fs* irradiation for fluences that are far below the ablation threshold for diamond. The latter, evidently, provide insufficient energy to convert significant volume/mass of diamond to graphite to attain any practical sample processing efficiency. As the fluence increases to 6.8 J/cm$^2$ the *I(D)/I(G)* ratio almost doubles to *ca.* 1.6, whereas the $L_a$ is reduced by ~50% (Fig. 2a) down to 10 nm, which corresponds to a further increase of *sp$^2$* aromatic clustering, and the reduction of *sp²* crystallinity in the 30 *fs*-irradiated regions of CVD samples.[23-24] Likewise, the $D_{HWHM}$ (Fig. 2b) increases from *ca.* 140 cm$^{-1}$ to *ca.* 175 cm$^{-1}$ with increasing fluence from 2.2 J/cm$^2$ to 3.9 J/cm$^2$ indicating an increase of structural disorder of *sp²* aromatic clusters[23, 33-34], followed by a unambiguous increase to over *ca.* 200 cm$^{-1}$ as the fluence is increased to 6.8 J/cm$^2$, confirming significant *sp²* amorphization[25] in the 30 *fs*-irradiated regions[27, 35].

High quality of *sp³* converted graphite-like clustering is observed up to 3.9 J/cm$^2$, except for the highest applied fluence of 6.8 J/cm$^2$, for which the $D_{HWHM}$ is sharply increased, indicating an increase in amorphization[25]. The $G_{HWHM}$ (Fig. 2b) values remain largely unchanged indicating that the *sp²* fractional disorder is marginally affected by the increasing fluence above the diamond ablation threshold. The D peak redshifts about 15 cm$^{-1}$ (Fig. 2c) as the fluence is increased from 2.2 J/cm$^2$ to 3.9 J/cm$^2$, corresponding to the increase of *sp²* aromaticity[35] and, as the fluence is further increased to 6.8 J/cm$^2$ the D peak blueshifts *ca.* 20 cm$^{-1}$ (Fig. 3c) as an indication of increasing amorphization of the *sp²* fraction, in accordance with the earlier observations for $I(D)/I(G)$ ratio and the $D_{HWHM}$ and the $G_{HWHM}$ changes. The $G_\Gamma$ slightly blueshifts from 1590 cm$^{-1}$ to 1594 cm$^{-1}$ as the fluence is increased from 2.2 J/cm$^2$ to 6.8 J/cm$^2$. The $G_\Gamma$ position below 1600 cm$^{-1}$ as noted by Ferrari *et al.*[36] is attributed to an exclusive presence of aromatic rings as no olefinic chains exist in this vibrational region. This observation indicates that the 30-*fs* laser irradiated regions on CVD samples contain only *sp²* aromatic rings and the irradiated regions are completely devoid of any *sp²* olefinic fractions.

Optical (Fig. 3a-d) and SEM (Fig. 3e-h) micrographs show 30 *fs*-laser induced graphitized tracks on CVD samples at varying fluence (colour-coded Raman spectra shown in Fig. 1). Optical micrographs (Fig. 3a-d) display more visually contrasting and homogeneous *sp²* areas compared to the SEM (Fig. 3e-h) images; the latter show a pronounced darker centre ablation line within the main graphitized area with a lighter background, which is corresponding to the Gaussian intensity distribution of the 30-*fs* source beam. Notably, the ablation within the graphitised tracks is visible even at lowest test fluence of 2.2 J/cm$^2$ (Fig. 3a, 3e) using the 30 *fs*-light, well below the nominal CVD ablation threshold fluence (*i.e.,* above 3 J/cm$^2$) reported for *fs* pulse widths exceeding 100 *fs*[18] as shorter pulse duration increases the beam intensity, which reduces the ablation threshold.[7] No cracking and/or thermal damage was observed inside the *sp$^2$* graphitised areas even at the highest test fluence (*i.e.,* 6.8 J/cm$^2$).

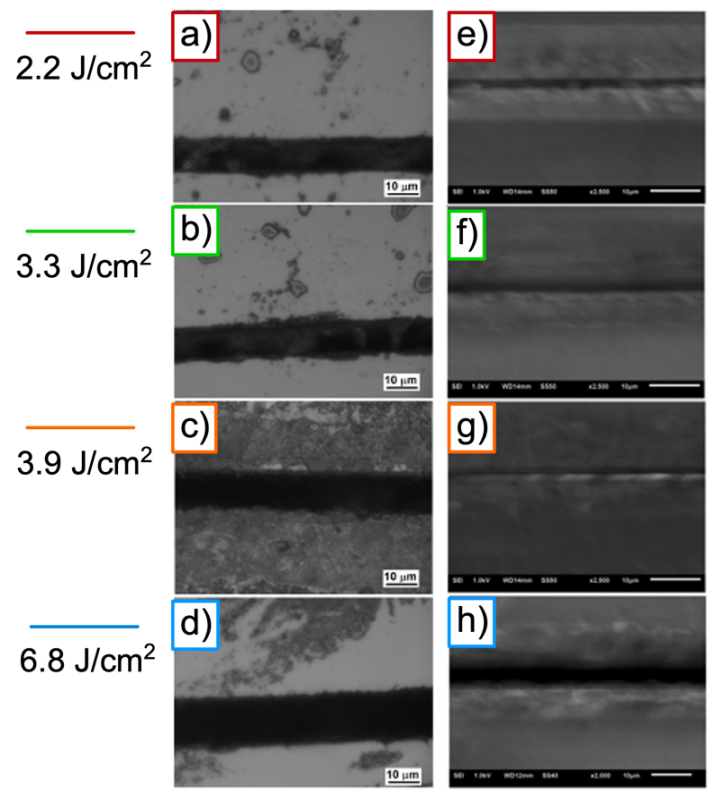

Fig. 3. Optical (a-d) and SEM (e-h) micrographs of 30 *fs*-irradiated regions on CVD samples at varying fluence (colour-coded to Raman spectra shown in Fig. 1); scale bar 10 $\mu$m.

The width, *w*, of the graphitised tracks was estimated from the optical images (Fig. 3(a-d)) in order to roughly determine the *sp$^2$* volume formation of the 30 *fs*-pulse irradiation process. The approximate track depth, *d* values were estimated from 514 nm Raman spectra by accounting for the reduced 1332 cm$^{-1}$ *sp$^3$* diamond core mode as suggested in the recent work by Komlenok *et. al.*[30] by applying the Beer-Lambert law, which relates the attenuation of light passing through a material to the properties of the material, as expressed in Eq. (2)

$$I(d) = I_0 e^{-\alpha d} \tag{2}$$

where, for diamond, $I$ is the 1332 cm$^{-1}$ diamond core mode intensity, $\alpha$ is the diamond optical absorption coefficient, and $d$ is depth of the *sp²* graphitized layer. The *sp²* layer absorbs both the probe beam and the scattering light and, therefore the intensity of the 1332 cm$^{-1}$ Raman line decreases owing to the reduced probe laser and the scattered signal intensities. Consequently, the $d$ value used to estimate the depth of the *sp²* graphitized layer, can be expressed as following:

$$d = \frac{\ln\left(\frac{I_{sp2}^{1332}}{I_{diamond}^{1332}}\right)}{2\alpha} \tag{3}$$

where, $I_{sp2}^{1332}$ and $I_{diamond}^{1332}$, are the intensities of 1332 cm$^{-1}$ core diamond Raman line taken from the graphitised *sp²* regions and the original (unmodified) diamond's surface, respectively. The $\alpha$ has been measured empirically and reported to be $1.16 \times 10^5$ cm$^{-1}$ for the laser induced *sp²* graphitized layer on diamond. [30]

The experimentally measured widths, *w* and the calculated depths, *d*, of 30 *fs*-laser induced graphitized tracks on CVD at varying fluence are shown in Fig. 4a. The *w* values were found to almost linearly increase with the increasing fluence in agreement with the earlier works on *fs*-laser irradiation of diamond[37], diamond-like carbon[38], dielectrics[39-40] and metals[41]. However, the estimated depth of the *sp²* tracks was found to be affected more markedly by the increase of the laser fluence, - such as the increase of fluence by approximately three (× 3) times resulted in an increased depths of the graphitized tracks over an order of magnitude (× 10) and, in an instance when the irradiation was performed near or at the ablation threshold, - a fractional increase of fluence (~20%) resulted in a three-fold increased depth within the $sp^2$ graphitised areas.

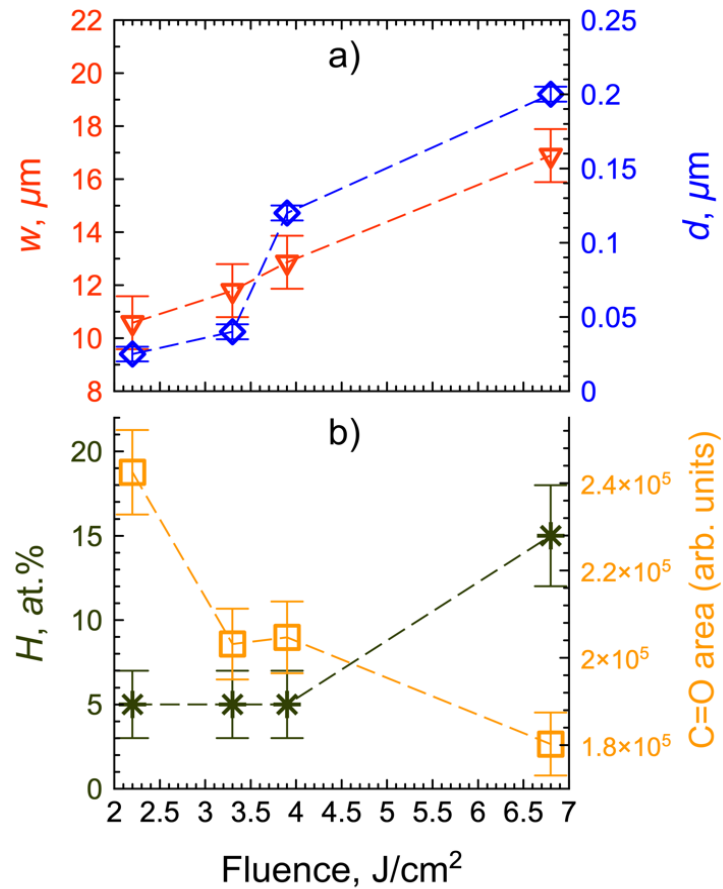

Fig. 4. a) As measured the effective widths, $w$, and the calculated depths, $d$, and b) the estimated hydrogen content, $H$, at.% and relative C=O contributions of 30 $fs$-laser induced graphitized tracks on CVD at varying laser fluence.

Notably, the $d$ values increased from a few tens of nanometres for a low fluence (*i.e.,* ca. 0.025 $\mu$m for the 2.2 J/cm$^2$) to a sub-micron scale at a high fluence (*i.e.,* 0.2 $\mu$m for 6.8 J/cm$^2$). At the highest fluence of 6.8 J/cm$^2$, no 1332 cm$^{-1}$ Raman line was detectable on $fs$-laser induced graphitised areas of the CVD samples. Such a non-linear effect of increasing fluence on the graphitization depths, we believe, is attributed to the non-linear optical absorption phenomena and multi-photon ionisation, which is a characteristic feature of the ultrashort *fs*-laser irradiation by which an excitation of electrons from the valence band into its conduction band is followed by a transient rapid change of the interatomic potential.[20] In the instance of our experiments, the ablation maxima are confined to the centre of the *sp²* graphitized tracks owing to the Gaussian intensity distribution of the beam that is consistent with the earlier findings by Abdelmalek *et al.*[42].

Owing to a pronounced PL background displayed in all Raman spectra (see Fig. 1) an approximate hydrogen (H-) content (at. %) in the irradiated *sp²* regions could be estimated using the work of Casiraghi *et. al.*[26] for green light excitation. It was estimated (Fig. 4b) that the *fs* laser-induced graphitised *sp²* regions generally displayed a 5 at.% protonation at or below 3.9 J/cm² fluence and a *ca.* 15 at.% for a higher 6.8 J/cm² fluence. Protonation in *fs*-laser irradiated regions is generally an indication of optical breakdown and the reduction of both the optical and Tauc gaps.[43,44] Naturally, a suppression and cleaving of C=O carbonyl group is observed with an increasing fluence in the irradiated *sp²* regions (Fig. 1a and Fig. 4b). Increasing shielding effects are expressed by a surface-localised plasma at CVD sample irradiation site as irradiation fluence is increased resulting in plasma reflecting and/or absorbing incoming *fs*-laser light photons[45-46], thus causing an extended shielding effect in the focal volume preventing an $O_2$ (from air) to reach the irradiation site. It is not known, however, how the degree of hydrogenation and the abundance of C=O group is distributed vertically, transversely and medially across the *sp²* photo- transformed regions in diamond. It is reasonable to assume that only the uppermost surface layer of a few nanometres thick is affected. It is, however, important to remember that in instances when the diamond *sp³* 1332 cm⁻¹ peak is detectable for visible excitations, the actual cross-sectional area for graphite at 514 nm is *ca.* 50 larger than that of $sp^3$ diamond phase as discovered by Wada et al.[47], and the hydrogenated amorphous carbon displays over a ~200 times larger cross-sectional area than diamond as noted by Salis and co-workers [48]. These observations suggest that the estimated depth of the 30 *fs*-light photo-induced $sp^2$ tracks on CVD is likely to be underestimated and the actual $sp^2$-rich affected regions are much larger. The current findings however, offer a valuable insight into the unique processes occurring on CVD under 50 *fs* laser light irradiation.

The following conclusions arise from our study:

- The application of an ultra-fast 30 *fs* laser irradiation to CVD samples up to a 3.9 J/cm$^2$ fluence was found to convert diamond's tetrahedral $sp^3$ phase into the $sp^2$ aromatic phase with a high degree of crystallinity without the formation of a HAZ or thermal cracking; above 3.9 J/cm$^2$ the size of planar crystalline domains in *fs* photo-irradiated regions is significantly reduced whereas the $sp^2$ aromatic clustering is preserved.

- *Fs*-irradiation of CVD near or at the ablation threshold was found to significantly influence the generated relative volume of the converted $sp^2$ phase, such as the increase of fluence by approximately three ($\times$ 3) times resulted in an increased depths of the graphitized tracks over an order of magnitude ($\times$ 10), and in an instance when the irradiation was performed near or at the ablation threshold, - a fractional increase of fluence (~20%, from 3.3 J/cm$^2$ to 3.9 J/cm$^2$) resulted in a three-fold increased depth of the $sp^2$ graphitised areas;

- Our results indicate that the application of CVD using sub-50 *fs* pulses (i.e., 30 *fs*) is associated with a significantly reduced (by approximately 20 – 30%) CVD ablation threshold values comparing to the instances where long(er) *fs*-pulse widths (*i.e.,* 60 – 100 *fs*) are used;

- Increased (presumably, surface-bound) hydrogenation was observed within the $sp^2$ graphitised regions on CVD in instances where the *fs*-irradiation was performed above the CVD ablation threshold; no marked hydrogenation increase has been observed for 30 *fs*-ablation below or near the CVD ablation threshold. The C=O cleavage occurred gradually with the increasing of irradiation fluence of 30 *fs* laser light.

CONFLICT OF INTEREST

The authors have no conflict of interest to disclose.

AUTHOR'S CONTRIBUTIONS

BA conducted experiments, collected and analysed data and drafted the manuscript. HX, DC designed and built the experimental set up. RS revised the manuscript. IVL designed the study and revised the manuscript. MR designed the study, analysed data, drafted and critically revised the manuscript. All authors read, commented on the manuscript, gave the final approval for publication and agree to be held accountable for the work performed herein.